\documentstyle[amsfonts,preprint,pre,aps,eqsecnum]{revtex}
\begin{document}
\draft
\title{Wormholes and Black Hole Pair Creation}
\author{Remo Garattini\thanks{%
e-mail: Garattini@mi.infn.it}}
\address{Facolt\`a di Ingegneria, Universit\`a di Bergamo\\
Viale Marconi, 5, 24044 Dalmine (Bergamo) Italy\\
BGTH/R-96/3}
\date{August 31, 1996}
\maketitle

\begin{abstract}
We analyze the possibility of black holes pair creation induced by three
dimensional wormholes. Although this spacetime configuration is nowadays
hard to suppose, it can be very important in the early universe, when the
wormhole spacetime foam representation can be meaningful. We compare our
approach with the no-boundary prescription of Hartle-Hawking.
\end{abstract}


\vfill
\eject

\section{Introduction}

\label{i}In recent years, most energies are being focussed on the study of
wormholes as significant contributions to quantum gravity, either in the
Euclidean or Minkowski formalism. Moreover, it has been argued that
wormholes may affect the constant of nature: in particular, they may lead to
a vanishing cosmological constant. On the other hand, wormholes can be used
as probes for studying the interior of a black hole and finally, an
important class termed traversable may well be included to remark the
importance of the subject in its various aspects. A classical wormhole is
defined by a throat of a given radius, connecting two asymptotically flat
three-dimensional spaces, an example can be the Schwarzschild metric, 
\begin{equation}
ds_{{\it 4}}^2=-\left( 1-\frac{2MG}r\right) dt^2+\left( 1-\frac{2MG}r\right)
^{-1}dr^2+r^2d\Omega ^2,  \label{i2}
\end{equation}
describing a four-dimensional wormhole with the horizon located at $2MG$. $M$
is the wormhole mass, $G$ is the gravitational constant and $d\Omega
^2=d\theta ^2+\sin ^2\theta d\phi ^2$. Another interesting wormhole is
described by the Schwarzschild-de Sitter metric 
\begin{equation}
ds_{{\it 4}}^2=-\left( 1-\frac{2MG}r-\frac \Lambda 3r^2\right) dt^2+\left( 1-%
\frac{2MG}r-\frac \Lambda 3r^2\right) ^{-1}dr^2+r^2d\Omega ^2.  \label{i5}
\end{equation}
When $\Lambda =0$, then the metric describes $\left( \text{\ref{i2}}\right) $%
, while when $M=0$, we obtain the de Sitter metric. By examining the roots
of $\left( \text{\ref{i5}}\right) $, we can distinguish two horizons located
at 
\begin{equation}
\ r_{+}=\frac 1{\sqrt{\Lambda }}\cos \left( \frac \theta 3\right) ,\ r_{++}=%
\frac 1{\sqrt{\Lambda }}\cos \left( \frac \theta 3+\frac{4\pi }3\right) ,
\end{equation}
where 
\begin{equation}
\cos \theta =-3m\sqrt{\Lambda },
\end{equation}
with the condition 
\begin{equation}
0\leq 9m^2\Lambda \leq 1.
\end{equation}
Here$\ r_{+}$ is the coordinate relative to the black hole horizon and $\
r_{++}$ is relative to the cosmological horizon. If $9m^2\Lambda =1$ then $%
r_{++}=r_{+}$ and the singularity is eliminated. This case can be considered
as extreme, exactly like the extreme Reissner-Nordstr\"om metric. However,
since $9m^2\Lambda =1$ the cosmological horizon and the black hole horizon
have merged. To avoid merging of horizons, we refer to the Nariai metric
described in ref. \cite{Perry} 
\begin{equation}
ds_{{\it 4}}^2=-\left( 1-\Lambda p^2\right) dt^2+\left( 1-\Lambda p^2\right)
^{-1}dp^2+\frac 1\Lambda d\Omega ^2.  \label{i9}
\end{equation}
The choice of these particular metrics is not casual. Indeed, when we
consider the Euclidean sections of $\left( \text{\ref{i2}}\right) $ and $%
\left( \text{\ref{i5}}\right) $, we discover negative modes in the saddle
point approximation \cite{Gross,Perry,Young}, which means instability with
respect to the reference asymptotic background (i.e., flat space and de
Sitter space). Moreover, both topology and signature changes are involved
together with tunnelling. Then we can see that a rich variety of phenomena
appear as a consequence of negative modes. At this point, we can ask
ourselves if the same result can appear when the Lorentzian signature is
used. Before approaching this problem, an intermediate step has to be
investigated and precisely: what happens if we split the time and space in $%
3+1$ dimensions? Concerning this question, the answer comes from an analysis
in a Kaluza-Klein type space, where the extra-dimension is represented by a
compact manifold. On this type of product space, we discover instability 
\cite{Witten}. The basic strategy is: take a manifold whose topology is $%
M^4\times B$, where $M^4$ represents the four-dimensional Minkowski
spacetime, while $B$ is a compact manifold of given radius. Then the line
element can be written, after an analytic continuation to euclidean space,
as 
\begin{equation}
ds^2=dr^2+r^2d\Theta ^2+d\phi ^2,
\end{equation}
where $r$ runs from $0$ to $\infty $ and $d\Theta ^2$ is the line element of
the three sphere. Comparing this solution with the five-dimensional black
hole solution \cite{Tangherlini} 
\begin{equation}
ds^2=\frac{dr^2}{1-\left( \frac Rr\right) ^2}+r^2d\Theta ^2+\left( 1-\left( 
\frac Rr\right) ^2\right) d\phi ^2,  \label{sc1}
\end{equation}
we discover that they have the same asymptotic behaviour. Since $\phi $ is a
periodic variable with periodicity $2\pi R$, we have now a non-singular
space which asymptotically approaches the Kaluza-Klein vacuum. Presence of
negative action modes in small fluctuations around $\left( \ref{sc1}\right) $
showing instability of $M^4\times S^1$ is guaranteed by the analogous
problem treated in one less dimension \cite{Gross}. Motivated by this and by
recent results on black hole pair production in inflationary cosmology \cite
{Bousso-Hawking}, we will investigate the possibility of black hole pair
production induced by wormholes, in particular three-dimensional wormholes,
which are nothing but ``{\it sections }'' of four dimensional metrics, with $%
t=const$. Such sections are {\it holes} in space but not in time. Recalling
that the probability of decay per unit time per unit volume is given by 
\begin{equation}
P\sim \left| \exp \left( -I\right) \right| ^2\sim \left| e^{-\left( \Delta
E\right) \left( \Delta t\right) }\right| ^2,  \label{pr}
\end{equation}
it is immediate to see that a possible approach to black hole pair
production is by means of the ``{\it Energy ''.} Nevertheless, this choice
needs a careful statement because constraint equations both at classical and
quantum level impose that dynamics is frozen. Indeed, if we look at the
quantum constraint equation\cite{DeWitt} 
\begin{equation}
{\cal H}\Psi =0,  \label{i1}
\end{equation}
known as Wheeler-DeWitt equation $\left( \text{WDW}\right) ,$ we see that it
represents an eigenvalue problem with zero eigenvalue. However eq. $\left( 
\text{\ref{pr}}\right) ,$ clearly indicates a non vanishing $\Delta E,$
otherwise even the action should be meaningless. This means that the problem
of computing energy is meaningless for, whatever configuration we fix, we
will obtain the same result. Two possibilities of solving this puzzling
question come into play. The first one is the resolution of the WDW equation
in superspace or in its reduced version of mini-superspace. The second one
is the resolution of the algebra of constraints generated by ${\cal H}$ and $%
{\cal H}_i$. Although this is a well tested way for probing quantum aspects
of gravity it leaves the question of the energy an open issue. Nevertheless,
if quantum aspects enter in gravity with some basic principles, they have to
be related to the possibility of having a functional Schr\"odinger equation.
Because of $\left( \text{\ref{i1}}\right) $, such an equation cannot exist
and the associated eigenvalue problem cannot be taken under consideration.
However, following ref. \cite{Remo1}, we shall assume the validity of an ``%
{\it internal }'' Schr\"odinger equation associated to each lapse function
in such a way to recover an eigenvalue equation. The plan of the paper is
the following: in section \ref{p1}, we will examine the topology of the
problem, in section \ref{p2}, we give a simple comparison of the action
calculated in a covariant way and in its splitted form of three space and
one time; in section \ref{p3}, we illustrate the results obtained in a
saddle point approximation both in four dimensions and in three plus one
dimensions. For this purpose we shall use a gaussian variational approach
which is very close to the quadratic approximation; in section \ref{p4}, we
approach the black hole pair production problem by comparing the results of 
\cite{Bousso-Hawking} with our approach; in section \ref{p5}, we summarize
and conclude.

\section{The Topology of the Problem}

\label{p1}In ref. \cite{Witten}, it was shown that, not only the compact
Euclidean periodically identified flat four dimensional manifold was
unstable, but even its $\left( 3+1\right) $ dimensional version. This
suggests the possibility of some basic relations between the
three-dimensional space and the four-dimensional one. Here we summarize some
aspects of the different topologies associated with metrics $\left( \text{%
\ref{i2}}\right) $ and $\left( \text{\ref{i5}}\right) $displaying whenever
possible the underlying three-dimensional structure.

\begin{description}
\item[a)]  Let us consider first the static, spherically symmetric metric of
a three dimensional wormhole 
\begin{equation}
ds_{{\it 3}}^2=\left( 1-\frac{2M}r\right) ^{-1}dr^2+r^2d\Omega ^2,
\label{a1}
\end{equation}
whose topology is ${\Bbb R\times S}^2.$ Metric $\left( \text{\ref{a1}}%
\right) $ possesses an apparent (horizon) singularity at $r=2M$, that is
removable by a suitable coordinate transformation. By defining 
\begin{equation}
dx=\pm \frac{dr}{\sqrt{1-\frac{2M}r}},  \label{aa1}
\end{equation}
metric $\left( \text{\ref{a1}}\right) $ is transformed into\footnote{%
This change of coordinates is present in the mathematics of embedding, i.e.,
when one wishes to construct, in three-dimensional Euclidean space, a
two-dimensional with the same geometry at the slice with {\it t }fixed.
Formula $\left( \text{\ref{aa1}}\right) $, defines the proper radial
distance as measured by static observers. The positive sign means that we
are performing calculations in the {\it upper universe.}} 
\begin{equation}
ds^2=dx^2+r^2\left( x\right) d\Omega ^2,  \label{aa2}
\end{equation}
where $x\in \left[ 0,\infty \right) .$ However, metric $\left( \text{\ref{a1}%
}\right) $ can be regarded as a section of a four dimensional metric. In
particular, if we consider constant time sections, i.e. $t=const$, we obtain
the Schwarzschild wormhole $\left( \text{\ref{i2}}\right) $ and if we
consider constant Euclidean time $\tau =-it$ sections we obtain the
Gibbons-Hawking instanton. To avoid conical singularities, the imaginary
time axis is periodic with period $\beta =8\pi M$. Then two routes for
enlarging the topology are possible. The first one is by choosing the
Euclidean metric with topology ${\Bbb R\times S}^2{\Bbb \times S}^1$ and the
second one is by choosing the Lorentzian metric ${\Bbb R}^2{\Bbb \times S}^2$%
. In any case the underlying wormhole topology is preserved and for the
Schwarzschild sector, we can summarize in the following diagram the topology
enlargement 
\begin{equation}
\ 
\begin{tabular}{ccccc}
${\Bbb R}^2{\Bbb \times S}^2$ &  & $\longleftrightarrow $ &  & ${\Bbb %
R\times S}^2{\Bbb \times S}^1$ \\ 
Lorentzian & $\nwarrow $ &  & $\nearrow $ & Euclidean \\ 
&  & ${\Bbb R\times S}^2$ &  & 
\end{tabular}
\label{a2}
\end{equation}

\item[b)]  On the other hand the time section of the Schwarzschild-de Sitter
metric is 
\begin{equation}
ds_{{\it 3}}^2=\left( 1-\frac{2M}r-\frac 13\Lambda r^2\right)
^{-1}dr^2+r^2d\Omega ^2.  \label{a5}
\end{equation}
The Schwarzschild-de Sitter metric has the wormhole topology, but the
interesting case is the extremal case $\left( \text{\ref{i9}}\right) $.
After eliminating the singularities, by analogy we continue to use the
(improper) term of wormhole also for the constant ``{\it time'' }section,
whose form metric is 
\begin{equation}
ds_3^2=\left( 1-\Lambda p^2\right) ^{-1}dp^2+\frac 1\Lambda d\Omega ^2.
\label{a9}
\end{equation}
Similarly to the Schwarzschild sector we can consider the diagram 
\begin{equation}
\begin{tabular}{ccccc}
${\Bbb H}^2{\Bbb \times S}^2$ &  & $\longleftrightarrow $ &  & ${\Bbb S}^2%
{\Bbb \times S}^2$ \\ 
Lorentzian & $\nwarrow $ &  & $\nearrow $ & Euclidean \\ 
&  & ${\Bbb S}^1{\Bbb \times S}^2$ &  & 
\end{tabular}
.  \label{a10}
\end{equation}
representing the possibilities of completing the topology.
\end{description}

\section{Computing the Action on compact topologies}

\label{p2}

As a simple application, we calculate here the action contribution on some ``%
{\it test}'' topologies, comparing the results obtained directly in {\it 4D}
with the ones obtained in {\it 3+1} dimensions. To obtain finite action
contributions, we fix our attention on some compact gravitational
instantons, in particular:

\begin{itemize}
\item  ${\Bbb S}^4$, correponding to the compact de Sitter metric (de
Sitter),

\item  ${\Bbb S}^2\times {\Bbb S}^2$, corresponding to the Nariai metric,
that is the extreme Schwarzschild-de Sitter metric,

\item  ${\Bbb S}^2\times {\Bbb S}^1\times {\Bbb R}$, corresponding to the
Schwarzschild metric.
\end{itemize}

\subsection{The action in $4$D}

\label{p1a}

\begin{description}
\item[a)]  ${\Bbb S}^4$

The action is represented by 
\begin{equation}
I=-\frac 1{16\pi G}\int\limits_{{\cal M}}d^4x\sqrt{g}\left( R-2\Lambda
\right) ,  \label{b1}
\end{equation}
where $\Lambda $ is the positive cosmological constant. By the Einstein
equations 
\begin{equation}
R_{\mu \nu }-\frac 12g_{\mu \nu }R+\Lambda g_{\mu \nu }=0\Longrightarrow
R=4\Lambda .  \label{b2}
\end{equation}
and $\left( \text{\ref{b1}}\right) $ becomes 
\begin{equation}
I=-\frac{2\Lambda }{16\pi G}\int\limits_{{\cal M}}d^4x\sqrt{g}=-\frac{%
2\Lambda }{16\pi G}V_{{\cal M}}^{\left( 4\right) },  \label{b3}
\end{equation}
where, $V_{{\cal M}}^{\left( 4\right) }=\frac 83\pi ^2\left( \sqrt{\frac 3%
\Lambda }\right) ^4$ is the four dimensional volume of the compact manifold.
Then 
\begin{equation}
I=-\frac{2\Lambda }{16\pi G}\frac 83\pi ^2\left( \sqrt{\frac 3\Lambda }%
\right) ^4=-\frac{3\pi }{G\Lambda }.  \label{b4}
\end{equation}

\item[b)]  ${\Bbb S}^2\times {\Bbb S}^2$

This is the topological product of two spheres of radius $\sqrt{\frac 3%
\Lambda }.$ The action is represented by eq. $\left( \text{\ref{b3}}\right) $
and $V_{{\cal M}}^{\left( 4\right) }=16\pi ^2\left( \sqrt{\frac 1\Lambda }%
\right) ^4$ and the final value of the action is 
\begin{equation}
I=-\frac{2\Lambda }{16\pi G}16\pi ^2\left( \sqrt{\frac 1\Lambda }\right) ^4=-%
\frac{2\pi }{G\Lambda }.  \label{b5}
\end{equation}

\item[c)]  ${\Bbb S}^2\times {\Bbb S}^1\times {\Bbb R}$

In this case the action needs a boundary term, otherwise the path integral
is meaningless. Here $\Lambda =0$ and $I$ is 
\begin{equation}
-\frac 1{16\pi G}\int\limits_{{\cal M}}d^4x\sqrt{g}R-\frac 1{8\pi G}%
\int\limits_{{\cal \partial M}}d^3x\sqrt{h}\left[ K\right] ,  \label{b6}
\end{equation}
where $\left[ K\right] $ is the difference in the trace of the second
fundamental form of ${\cal \partial M}$ in the metric $g$ and the metric $%
\eta $ referred to the flat space. In this case the Einstein equation give $%
R_{\mu \nu }=0$. This means $R=0$ and the action is given by 
\begin{equation}
I=-\frac 1{8\pi G}\int\limits_{{\cal \partial M}}d^3x\sqrt{h}\left[ K\right]
=4\pi GM^2,  \label{b7}
\end{equation}
where we have used the fact that the Euclidean ``time'' is periodic with
period $8\pi GM$ and the fact that the hypersurface is bounded by the
surface $r=r_0$.
\end{description}

We will see in next section that all these contributions for the different
topologies will be extracted by the same hypersurface term giving the
relation between $4D$ and $\left( 3+1\right) D$.

\subsection{The Action in $\left( 3+1\right) $D}

\label{p1b}

The scalar curvature appearing in the action described in terms of {\it lapse%
} and {\it shift} variables is 
\begin{equation}
^{\left( 4\right) }R=^{\left( 3\right) }R-K_{ij}K^{ij}+K^2-\frac 2N\frac{%
\partial K}{\partial \tau }+\frac 2N\left[ \nabla ^2N+N^p\left( K\right)
_{|p}\right] ,  \label{c1}
\end{equation}
where $K$ is the trace of the second fundamental form and $\tau $ is the
Euclidean time related by $t=-i\tau $. Since we consider only ``static
spherically symmetric'' metrics, the term involving $N^p$ disappears and $%
K_{ij}=0$, therefore 
\begin{equation}
^{\left( 4\right) }R=^{\left( 3\right) }R+\frac 2N\nabla ^2N,  \label{c2}
\end{equation}
that is, the contribution for these metrics comes only from a boundary term
and from the three scalar curvature. This is a consequence of the fact that
we have opened the hypersurfaces, i.e. they do not represent more a compact
object. For example, the de Sitter metric can be represented by 
\begin{equation}
ds^2=d\tau ^2+\cos ^2\tau d\Omega _3^2,  \label{c2a}
\end{equation}
which has a compact ${\Bbb S}^4$ topology whose spatial section is a closed
(compact) ${\Bbb S}^3$ hypersurface, otherwise we can choose the ``{\it %
static''} representation 
\begin{equation}
ds^2=\left( 1-\frac \Lambda 3r^2\right) d\tau ^2+\left( 1-\frac \Lambda 3%
r^2\right) ^{-1}dr^2+r^2d\Omega ^2,  \label{c2b}
\end{equation}
where for every choice of $\tau =const$, the hypersurface covers the region
inside the cosmological horizon, giving therefore open topologies\footnote{%
The same procedure is used in ref. \cite{Englert} to the approach of entropy
generation by a quantum tunnelling.}\cite{Tanaka}. For this reason, we see
immediately that 
\begin{equation}
^{\left( 4\right) }R=+\frac 2N\nabla ^2N.  \label{c3}
\end{equation}
In fact $^{\left( 3\right) }R=0$, for the Schwarzschild metric and since we
have to consider $^{\left( 4\right) }R-2\Lambda $ for de Sitter and
Schwarzschild-de Sitter (Nariai) metrics $^{\left( 3\right) }R=2\Lambda $,
we have the result of eq.$\left( \text{\ref{c3}}\right) .$ The action is
represented by the surface term 
\begin{equation}
I=-\frac 1{16\pi G}\int d\tau \int d^3xN\sqrt{^3g}\left[ \frac 2N\nabla
^2N\right] =-\frac 1{8\pi G}\int d\tau \int d^3x\left[ \partial _i\left( 
\sqrt{^3g}g^{ij}\partial _jN\right) \right] .  \label{c4}
\end{equation}
Now we verify that eq.$\left( \text{\ref{c4}}\right) $ matches with the
results of the previous section.

\begin{description}
\item[a)]  de Sitter vs. ${\Bbb S}^4$ topology.

The line element is 
\begin{equation}
ds^2=\left( 1-\frac \Lambda 3r^2\right) d\tau ^2+\left( 1-\frac \Lambda 3%
r^2\right) ^{-1}dr^2+r^2d\Omega ^2,  \label{c5}
\end{equation}
and identifying $N^2=1-\frac \Lambda 3r^2$, then $\left( \text{\ref{c4}}%
\right) $ becomes 
\[
I=-\frac 1{8\pi G}\int d\tau \int d^3x\left[ \partial _i\left( \sqrt{^3g}%
g^{ij}\partial _jN\right) \right] =-\frac \Lambda {6G}\int d\tau r^3|_{=%
\sqrt{\frac 3\Lambda }}=-\frac \Lambda {6G}\left( 2\pi \sqrt{\frac 3\Lambda }%
\right) \left( \frac 3\Lambda \right) ^{\frac 32} 
\]
\begin{equation}
=-\frac{3\pi }{\Lambda G}.  \label{c6}
\end{equation}

\item[b)]  Nariai vs. ${\Bbb S}^2\times $ ${\Bbb S}^2$ topology.

The line element is 
\begin{equation}
ds^2=\left( 1-\Lambda r^2\right) d\tau ^2+\left( 1-\Lambda r^2\right)
^{-1}dr^2+\frac 1\Lambda d\Omega ^2,  \label{c7}
\end{equation}
and performing the same calculation for the a) case we obtain 
\begin{equation}
I=-\frac 1{8\pi G}\left( \frac{2\pi }{\sqrt{\Lambda }}\right) \left( 4\pi 
\frac 1\Lambda \right) \int d\tau =-\frac{2\pi }{G\Lambda }.  \label{c8}
\end{equation}

\item[c)]  Schwarzschild vs. ${\Bbb S}^2\times $ ${\Bbb S}^1\times {\Bbb R}$
topology.

The line element is 
\begin{equation}
ds_{}^2=\left( 1-\frac{2M}r\right) d\tau ^2+\left( 1-\frac{2M}r\right)
^{-1}dr^2+r^2d\Omega ^2,  \label{c9}
\end{equation}
and the action becomes 
\[
I=-\frac 1{8\pi G}4\pi \left[ \int dr\partial _r\left( r^2\sqrt{1-\frac{2MG}r%
}\partial _r\sqrt{1-\frac{2MG}r}\right) \right] _{r=r_0}\int d\tau =-\frac 1{%
8\pi G}4\pi \left( MG\right) \left( 8\pi MG\right) 
\]
\begin{equation}
=\frac 1{8\pi G}\int\limits_{{\cal \partial M}}d^3x\sqrt{h}\left[ K\right]
=-4\pi M^2G.  \label{c10}
\end{equation}
As shown, this gives exactly the term $\left( \text{\ref{b7}}\right) $ with
the reversed sign.
\end{description}

\section{Negative modes in $4$D and in $\left( 3+1\right) $D}

\label{p3} In previous sections we examined the action contribution of some
compact topologies from a general point of view. In particular, we have
verified that by opening the topologies the exponential background factor
agrees with the value obtained performing the calculation on such compact
objects. Moreover, the non vanishing surface term that will be related to
the decay exponential is deduced from the same term, i.e. the horizon
hypersurface. Here, we report, how the action and consequently the partition
function modify when a fixed background choice is made to estimate decay
ratios. If $g=\overline{g}+h$, then 
\begin{equation}
I\left[ g\right] =I\left[ \overline{g}\right] +I_2\left[ \overline{g}%
,h\right] +\ldots \ldots ,  \label{d1}
\end{equation}
where $I_2$ is quadratic in the perturbations $\overline{g}$. In this
approximation the decay rate per unit volume is given by 
\begin{equation}
\Gamma =\left| A\right| \exp \left( -I\left[ \overline{g}\right] \right) ,
\label{d2}
\end{equation}
where $A$ is the prefactor coming from the evaluation of the one loop
expansion of the partition function. In particular, it is proportional to
its imaginary part.

\subsection{$4$D}

Limiting us to the transverse-traceless sector (TT or spin 2), eq. $\left( 
\text{\ref{d2}}\right) $ can be approximated by 
\begin{equation}
\Gamma =\left| A\right| \exp \left( -I\left[ \overline{g}\right] \right)
\simeq \left( \det M\right) ^{-\frac 12}\exp \left( -I\left[ \overline{g}%
\right] \right) ,  \label{d2a}
\end{equation}
where the determinant is represented by the usual gaussian formula 
\begin{equation}
\left( \det M\right) ^{-\frac 12}=\int {\cal D}h\exp -\frac 1{16\pi G}\int
d^4x\left[ \frac 14h_{ab}^{TT}M^{abcd}h_{cd}^{TT}\right] ,  \label{d3}
\end{equation}
and 
\begin{equation}
M^{abcd}h_{cd}^{}=-\Box h^{ab}-2R^{acbd}h_{cd}  \label{eu1}
\end{equation}
The eigenvalues coming from the eigenvalue equation associated with $\left( 
\text{\ref{eu1}}\right) $ are\footnote{%
We omit here the discussion on zero modes, since for our purposes this is
not relevant.}:

\begin{description}
\item[i)]  $\lambda =-2\Lambda $ for the degenerate Schwarzschild-de Sitter
case, i.e., the Nariai metric,\cite{Perry,Young}

\item[ii)]  $\lambda \simeq -0.19\left( GM\right) ^{-2}$ for the
Schwarzschild metric\footnote{%
However, B.\ Allen showed that the negative mode can be avoided by
restricting the box containing the black hole in such a way to forbid its
formation and its growing \cite{Allen}.}\cite{Gross}.
\end{description}

\subsection{$\left( 3+1\right) $D}

Following ref.\cite{Remo}, we expand the scalar curvature around a fixed
three dimensional background up to second order, then after having separated
the degrees of freedom by means of ultralocal metrics we fix our attention
on the resulting traceless transverse sector. The trial wave functional is
defined by 
\begin{equation}
\Psi _{\alpha }\left[ h_{ij}^{{}}\left( \overrightarrow{x}\right) \right] =%
{\cal N}\exp \left\{ -\frac{1}{4l_{p}^{2}}\left[ \left\langle
hK_{{}}^{-1}h\right\rangle _{x,y}^{\bot }\right] \right\} .
\end{equation}
$\left\langle \cdot ,\cdot \right\rangle _{x,y}^{{}}$ denotes space
integration and $K_{{}}^{-1}$ is the inverse propagator defined by 
\begin{equation}
K_{{}}^{\bot }\left( x,x\right) _{iakl}:=\sum_{N}\frac{h_{ia}^{\bot }\left(
x\right) h_{kl}^{\bot }\left( y\right) }{2\lambda _{N}\left( p\right) },
\end{equation}
where $\lambda _{N}\left( p\right) $ are infinite variational parameters and 
$h_{ia}^{\bot }\left( x\right) $ are the eigenfunctions of 
\begin{equation}
\left( -\triangle \delta _{j}^{a_{{}}^{{}}}+2R_{j}^{a}\right)
h_{a}^{i}=-E^{2}h_{j}^{i_{{}}^{{}}}.
\end{equation}
$\triangle $ is the Laplacian in a curved background and $R_{j}^{a}$ is the
mixed Ricci tensor. When the cosmological constant is considered, we have 
\begin{equation}
\left( -\triangle \delta _{j}^{a_{{}}^{{}}}+2R_{j}^{a}-2\Lambda \delta
_{j}^{a_{{}}^{{}}}\right) h_{a}^{i}=-E^{2}h_{j}^{i}.
\end{equation}
The ``-'' sign appears because we are searching for bound states of the two
previous equation. Indeed finding bound states is equivalent to searching
for negative modes. Solving the previous equations we obtain the following
eigenvalues:

\begin{description}
\item[i)]  $E^2\simeq -2\Lambda $ for the Nariai ``wormhole\thinspace ''case

\item[ii)]  $E^2\simeq -0.24\left( MG\right) ^{-2}$ for the Schwarzschild
wormhole.
\end{description}

The numerical discrepancy between the four dimensional case and the three
dimensional one is due to different numerical factors coming into play when
the Laplacian operator and the connections are computed.

\section{Black hole pair creation}

\label{p4}As mentioned in the introduction and related in section \ref{p3},
negative modes give evidence of instability. This means that these saddle
points are {\it bounces} instead of being instantons, that is at one loop
the approximated functional ceases to be gaussian. Since these backgrounds
are periodic in the imaginary time, this periodicity is usually interpreted
as finite temperature $T$. Then the instability of these configurations
happens for the hot space. {\it Gross et al. }interpreted this fact as
corresponding to a semiclassical {\it nonperturbative} instability of hot
flat space in a thermal bath due to the nucleation of black holes. The
appearance of a single negative eigenvalue is related to the bounce which
shifts the energy of the false ground state. Then neglecting the prefactor 
\cite{Coleman}, the approximate value of the probability of nucleating a
black hole per unit volume per unit time, which we denote as $\Gamma $, for
the Gibbons-Hawking instanton is 
\begin{equation}
\Gamma \sim \exp -4\pi M^2G,  \label{eu6}
\end{equation}
while for the Nariai instanton is 
\begin{equation}
\Gamma \sim \exp -\frac \pi {\Lambda G}.  \label{eu7}
\end{equation}
In any case the ``{\it hot}'' space cannot describe the ground state,
therefore a topology change comes into play and a black hole nucleation can
be realized when we consider the {\it hot flat space}, while black hole pair
creation is the mechanism related to the {\it hot de Sitter space. }It is
immediate to recognize that eq. $\left( \text{\ref{eu7}}\right) $ represents
the decay rate calculated with the no-boundary prescription of
Hartle-Hawking. In fact, following ref. \cite{Bousso-Hawking}, we define 
\begin{equation}
\Gamma =\frac{P_{\text{SdS}}}{P_{\text{de Sitter}}}=\exp -\frac \pi {%
G\Lambda }.  \label{de}
\end{equation}
On the other hand, in ref. \cite{Remo1} the choice for the wave function
solving the Wheeler-DeWitt equation was 
\begin{equation}
\Psi \left[ g_{ij}\right] =\int\limits_\gamma dN\ e^{-N\omega }\Phi \left[
g_{ij}\right] .  \label{wf}
\end{equation}
This form was obtained with the Euclidean action defined by\footnote{%
Actually, in ref. \cite{Remo1}, the exponential was imaginary because the
signature was of the Lontzian type.}\footnote{%
Obviously for the de Sitter and Nariai space, we have to correct the scalar
curvature by adding $-2\Lambda .$} 
\begin{equation}
I=-\frac 1{16\pi G}\int\limits_{{\cal M}}d^4x\sqrt{g}R-\frac 1{8\pi G}%
\int\limits_{{\cal \partial M}}d^3x\sqrt{h}K,
\end{equation}
however the three space plus one time integral over decomposition $\left( 
\text{\ref{c1}}\right) $represents the same previous action plus the
boundary at $\tau =0:$ i.e. eq. $\left( 2.4\right) $ of ref. \cite
{Bousso-Hawking}. Then eq. $\left( \text{\ref{wf}}\right) $ has to be
modified with 
\begin{equation}
\Psi \left[ g_{ij}\right] =\exp \left( \frac 1{8\pi }\int_{\tau =0}d^3x\sqrt{%
^3g}K\right) \int\limits_\gamma dN\ e^{-N\omega }\Phi \left[ g_{ij}\right] .
\end{equation}
Heuristically speaking, we can conjecture that the only difference with the
Hartle-Hawking wave function is that $\Phi \left[ g_{ij}\right] $ is a trial
wave functional of the gaussian type \cite{Remo}. It is quite obvious, at
this point, that by repeating the same procedure of cutting half of the
instanton, we recover the results that lead to eq. $\left( \text{\ref{de}}%
\right) .$ The same approach has to be applied for the Gibbons-Hawking
instanton which will be cutted in half following the same procedure of ref.%
\cite{Frolov}, where a no-boundary proposal for black holes was presented.

\section{Conclusions and outlooks}

\label{p5}Motivated by the recent results on black hole pair creation
obtained in the Euclidean sector that we have summarized in section \ref{p4}%
, we have tried to reproduce the same calculation scheme, but from the
Wheeler-DeWitt equation point of view. Although the choice of separating
time from space is dense of subtleties and technical problems, a partial
reproduction of the above mentioned results has been obtained. Our approach
has the same features of a Kaluza Klein space, if the {\it lapse} boundary
conditions are of periodic type, that it means that the spacetime has
topology $R^3\times S^1$ namely, flat spacetime is ``{\it hot}''. However,
although we limited our investigation to the TT sector and an improvement
including the spin 1 and spin 0 sector is necessary to give robust
conclusions, we can realize that our results about instabilities and pair
creation seem to suggest a parallel method for semiclassical evaluations, at
least for static metrics and consequently for a certain kind of topologies.
An interesting feature of this approach is that we can have a better control
on the signatures giving therefore hints on the understanding of some
evolution or thermal processes. In fact, coming back to the black hole pair
creation again, we see that due to boundary conditions, this process is a
consequence of a {\it hot} background space, namely the cold space is not
able in producing black holes or topology changes and therefore this cannot
be considered a {\it real} vacuum fluctuation. However, our approach seem to
be valid without imposing any periodicity and therefore any temperature.
This is a direct consequence of choosing a three-dimensional {\it wormhole,}
whose perturbations reside in the three-space, and since we have
investigated a stationary {\it energy levels} problem like in ordinary
Quantum Mechanics. An open problem, coming from this observation, could be
the possible decay of Minkowski space, namely the {\it zero temperature}
version of flat space. The positive energy conjecture forbids a classical
decay because $E=0$ if and only if we consider Minkowski space, otherwise $%
E\geq 0.$ However, as argued in ref. \cite{PGDiaz}, a possibility is related
to pairs of black hole that do not reside in the same universe, but they
propagate in universes separated by a wormhole. Actually, the best way to
treat this splitting of black holes is by introducing the concept of {\it %
quasilocal energy} \cite{Frolov1}. In our case, the Schwarzschild
three-dimensional wormhole is nothing but the spatial part of a static
Einstein-Rosen bridge (a {\it section} or a {\it slice}) and the quasilocal
energy can be written as 
\begin{equation}
E=E_{+}-E_{-}
\end{equation}
The total energy is zero for boundary conditions symmetric with respect to
the bifurcation surface $S_0,$ which is the case of the asymptotically flat
space, where the previous formula becomes 
\[
E_{Flat}=0\rightarrow M+\left( -M\right) +\text{ topology change.}
\]
$M$ represents the black hole (in the upper universe), while $\left(
-M\right) $ represents the {\it anti-black hole }(in the lower universe).
This is the same of cutting in half the four dimensional wormhole and
consider only one half on a fixed slice. In this situation, we have a non
vanishing probability of nucleating black holes, but in different {\it %
universes} \cite{Mazur,PGDiaz}. Although there is no direct evidence of the
existence of such wormholes in present days, they can be considered in the
very early universe, where according to ref. \cite{MTW}, pure gravitational
wormholes could be existed. This agrees with ref. \cite{Bousso-Hawking},
where neutral black hole pair creation could be relevant in the very early
universe, where the cosmological constant could assume the pre-inflationary
value $\Lambda \sim 1$ (in Planck's units). To this purpose, an effective
cosmological constant is needed and a way to do this is to include scalar
matter fields for driving the cosmological constant close to one Planck's
unit \cite{Bousso-Hawking}.To conclude, we want to remark that the black
hole pair creation could be considered a mechanism induced by
three-dimensional wormholes. Nevertheless, our analysis is limited to a
single wormhole, while it could be very interesting to consider a
multi-wormhole space. Such a collection of wormholes could be viewed as a
test for probing spacetime foam as a candidate for a quantum picture of
gravity.

\section{Acknowledgments}

I wish to thank H.D. Conradi, V. Moretti and P. Spindel for helpful
discussions.


\begin{references}
\bibitem{DeWitt}  B.S. DeWitt, Phys. Rev. {\bf 160} (1967), 1113.

\bibitem{Bousso-Hawking}  R. Bousso and S.W. Hawking, {\it The probability
for primordial black holes.} Phys. Rev. D {\bf 52} (1995), 5659 -
gr-qc/9506047; R. Bousso and S.W. Hawking, {\it Pair creation of black holes
during inflation.} Preprint no. DAMTP/R-96/33, gr-qc/9606052.

\bibitem{Perry}  P. Ginsparg and M.J. Perry, Nucl. Phys. B {\bf 222} (1983),
245.

\bibitem{Gross}  D.J. Gross, M.J. Perry and L.G. Yaffe, Phys. Rev. D{\bf \ 25%
} (1982), 330.

\bibitem{Remo}  R. Garattini, {\it Vacuum Energy in Ultralocal Metrics for
TT tensors with Gaussian Wave Functionals, }gr-qc/9508060.

\bibitem{Nariai}  S. Nariai, {\it On some static solutions to Einstein's
gravitational field equations in a spherically symmetric case. }Science
Reports of the Tohoku University, {\bf 34} (1950), 160; S. Nariai, {\it On a
new cosmological solution of Einstein's field equations of gravitation. }%
Science Reports of the Tohoku University, {\bf 35} (1951), 62.

\bibitem{MTW}  C.W. Misner, K.S. Thorne and J.A. Wheeler, {\it Gravitation}
(Freeman, San Francisco, 1973) 842; M.S. Morris and K.S. Thorne, Am. J. Phys.%
{\it \ }{\bf 56 }(1988), 395.

\bibitem{Halliwell}  J.J. Halliwell, ''Introductory Lectures on Quantum
Cosmology''. In {\it Jerusalem Winter School for Theoretical Physics:
Quantum Cosmology and Baby Universes Vol. 7}. S. Coleman, J.B. Hartle, T.
Piran and S. Weinberg, eds. World Scientific, 159-243.

\bibitem{Remo1}  R. Garattini, {\it Vacuum Energy Estimates in Quantum
Gravity and the Wheeler-DeWitt Equation, }gr-qc/9604004.

\bibitem{Mazur}  P.O. Mazur Mod. Phys. Lett. A {\bf 4}, (1989), 1497.

\bibitem{Young}  R.E. Young, Phys. Rev. D {\bf 28 (}1983), 2436; R.E. Young,
Phys. Rev. D {\bf 28} (1983), 2420.

\bibitem{Coleman}  S. Coleman, Nucl. Phys. B {\bf 298} (1988), 178.

\bibitem{Allen}  B. Allen, Phys. Rev. D {\bf 30} (1984), 1153.

\bibitem{Witten}  E. Witten, Nucl. Phys. B {\bf 195 (}1982), 481.

\bibitem{Tangherlini}  F.R. Tangherlini, Nuovo Cim. {\bf 27} (1963), 636.

\bibitem{Frolov}  A.O. Barvinsky, V.P. Frolov and A.I. Zelnikov, Phys. Rev. D%
{\bf \ 51} (1995), 1741.

\bibitem{PGDiaz}  P.F. Gonzalez-Diaz, {\it The Schwarzschild Black Hole Pair}%
. Preprint no. IMAFF-RC-04-95, gr-qc/9503011.

\bibitem{Frolov1}  V.P. Frolov and E.A. Martinez, {\it Action and
Hamiltonian for Eternal Black Holes, }Class.Quant.Grav.{\bf 13} (1996), 481,
gr-qc/9411001.

\bibitem{Englert}  A. Casher and F. Englert, Class.Quant.Grav. {\bf 9 }%
(1992), 2231.

\bibitem{Tanaka}  T. Tanaka and M. Sasaki, Prog. Theor. Phys. {\bf 88}
(1992), 503.
\end{references}
\end{document}